# Taylor Series Kinematics

Craig W. Looney
Merrimack College, North Andover, MA
(author email: looneyc@merrimack.edu)

Has it ever occurred to you that the kinematic equations for uniformly accelerated one-dimensional motion –

$$x = x_0 + v_0 t + \frac{1}{2} a t^2, \tag{1}$$

$$v = v_0 + at, \tag{2}$$

and

$$v^2 = v_0^2 + 2a(x - x_0) \tag{3}$$

– are Taylor series expansions? If not, you are in good company. I didn't know this myself until a colleague[1] pointed it out to me many years ago, and I was stunned to learn something new and wonderful about something so familiar. Accordingly, my first objective in this paper is to clearly present the not-widely-known Taylor series derivations of these basic equations to a population primed to deeply appreciate them: people, like me, who teach introductory physics. Following this, I use the Taylor series approach to derive a generalized one-dimensional expression for *x(t)* that includes the jerk and further kinematic time derivatives, which have importance in many real-world applications and in which there has been renewed pedagogical interest. I also outline teaching suggestions and provide student-accessible video derivations to support instructors who would like to incorporate Taylor series kinematics into their teaching, while identifying sequencing-related challenges. I close with the observation that the traditional second *calculus* course, which is largely free of sequencing issues, could be a great place to incorporate and leverage Taylor series kinematics, and I briefly outline an early-stage pilot collaboration to explore this possibility.

**Taylor series derivations of the constant acceleration kinematic equations**

The Taylor series expansion of a function $f(z)$ about $z = z_0$ is given by

$$f(z) = [f]_{z_0} + \left[\frac{df}{dz}\right]_{z_0}(z - z_0) + \frac{1}{2!}\left[\frac{d^2 f}{dz^2}\right]_{z_0}(z - z_0)^2 + \ldots + \frac{1}{n!}\left[\frac{d^n f}{dz^n}\right]_{z_0}(z - z_0)^n + \cdots \tag{4}$$

To derive Eq. (1), we first expand $x(t)$ about $t = t_0$:

$$x(t) = [x]_{t_0} + \left[\frac{dx}{dt}\right]_{t_0}(t - t_0) + \frac{1}{2!}\left[\frac{d^2 x}{dt^2}\right]_{t_0}(t - t_0)^2 + \ldots + \frac{1}{n!}\left[\frac{d^n x}{dt^n}\right]_{t_0}(t - t_0)^n + \cdots \tag{5}$$

The first and second time derivatives of $x$ are the velocity ($v$) and acceleration ($a$), respectively; in the case of constant acceleration, all higher time derivatives of $x$ vanish. Following the usual convention, we set $t_0 = 0$, and Eq. (1) immediately follows. A detailed video of the Taylor series derivation of Eq. (1), designed to be accessible to students, is publicly available in Ref. 2.

In a very similar way – the details are left to the reader – the expansion of $v(t)$ about $t = t_0$ leads directly to Eq. (2) in the case of constant acceleration. The Taylor series derivation of Eq. (3) involves expanding $v^2(x)$ about $x = x_0$; a detailed video derivation, designed to be accessible to students, is publicly available in Ref. 3. It is important to emphasize that while Taylor series expansions are often utilized as approximations, the Taylor series derivations of Eqs. (1), (2), and (3) are *not* approximations but rather are rigorous derivations of exact results in the special case of uniformly accelerated one-dimensional motion.

**Beyond constant acceleration: jerk, snap, crackle, pop …**

The Taylor series approach provides a nearly effortless opportunity to introduce generalized one-dimensional kinematics and higher order kinematic quantities on the way to deriving Eq. (1). The third time derivative of the position is commonly called the jerk ($j$), while snap ($s$), crackle ($c$), and pop ($p$) have emerged as consensus names for the 4th, 5th, and 6th time derivatives, respectively. With these definitions, and the assumption that $t_0 = 0$, Eq. (5) becomes

$$x = x_0 + v_0 t + \frac{1}{2!}a_0 t^2 + \frac{1}{3!}j_0 t^3 + \frac{1}{4!}s_0 t^4 + \frac{1}{5!}c_0 t^5 + \frac{1}{6!}p_0 t^6 + \cdots \qquad (6)$$

In my video derivation[2] of Eq. (1), introducing the higher order kinematic quantities and writing out Eq. (6) contributes less than 90 seconds to the total duration of the video. The corresponding generalized one-dimensional expression for $v(t)$ can be derived in a similar manner through the Taylor series approach, or by taking the time derivative of Eq. (6).

The jerk, despite vigorous interest and discussion over 30 years ago in the pages of *The Physics Teacher*[4,5,6,7,8,9,10], continues to be largely absent from university physics and engineering textbooks[11], unless we include the artificial infinite jerks that frequently appear in standard textbook problems. Contemporary[12,13,14,15,16] pedagogical interest – which, while not widespread, is on the rise – is motivated by the relevance of the jerk to variable-force interactions in our everyday experience and to a vast and ever-growing array of STEM applications. For example, railroad and highway transitions from straightaways to circular arcs have long been designed to limit[17,18] centripetal jerks by gradually rather than instantaneously decreasing the radius of curvature from infinity to that of the desired arc, often by using segments of the Cornu spiral – a.k.a. Euler spiral or clothoid – familiar to generations of optics students. Clothoid segments and other transition schemes are also commonly used in vertical roller-coaster loops[13,19] to produce teardrop-shaped rather than circular loops in order to smooth out the centripetal jerk upon entry and exit (Fig. 1). Refs. 12 and 13, written recently for an audience of physics instructors, consider the safety and experiential aspects of the jerk and snap for roller coasters[12,13] and trampolines[12] and include discussions of real accelerometer data. The preceding examples are but a small sample; an extensive catalog of further applications of the jerk, including but not limited to 3D printing, machine and motor control, seismic analysis, and greyhound racing, is available in a recent 2020 review article[11] written for STEM educators.

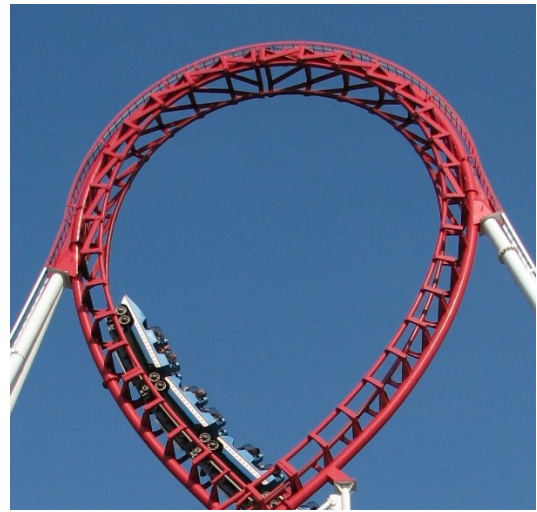

**Fig. 1.** The teardrop-shaped loop ensures a gradual rather than abrupt change in the roller-coaster's centripetal acceleration. Photo credit: Jeremy Thompson. Displayed photo is cropped from the original. Licensed under CC BY 2.0: https://creativecommons.org/licenses/by/2.0/deed.en

**Teaching suggestions and sequencing challenges**

I am *not* suggesting that Taylor series be used to *introduce* kinematics, but rather to make connections and expand context. The Taylor series derivation of Eq. (1) – with or without Eq. (6) en route – is one of the most straightforward physically relevant applications of Taylor series imaginable. In an alternative – but not necessarily desirable – reality in which all students taking introductory physics have already studied a full year of calculus, it would be natural to connect Taylor series to kinematics at the conclusion of a traditional treatment of kinematics. However, most introductory physics students do not encounter Taylor Series in their calculus courses until well after this otherwise natural opportunity has passed. Consequently, most students never see this connection, and apparently even most physics and mathematics instructors – myself included! – do not imagine this connection on their own. Nevertheless, for potentially interested instructors, alternative connection opportunities – necessarily contingent on local conditions, can be engineered. For example, Taylor series kinematics would make an excellent enrichment topic for a college or university recitation section or for the post-AP-exam phase of a high school AP Physics C course.

**Taylor series kinematics in the introductory calculus sequence**

In recent years I have come to the realization that the sequencing challenges noted above can be turned around, and leveraged as advantages, by teaching Taylor series kinematics in the introductory *calculus* sequence. Indeed, the overwhelming majority of students who study calculus long enough to encounter Taylor series have already learned about and used the constant acceleration kinematic equations in a prior college or high-school physics course. Furthermore, in college – and high school AP – calculus courses, kinematics is already a staple example application for differentiation and integration. Why not also lever the concrete physical context of kinematics to help motivate initial student interest in Taylor series, a topic that many students do not appreciate – or even imagine will ever be useful – until it is needed in subsequent coursework?

A mathematics colleague[20] with whom I had have been discussing this idea recently undertook an initial low-stakes pilot of Taylor series kinematics at the end of the sequences and series unit in her calculus 2 course. In-class activities focused on engaging students in deriving and discussing Eqs. (1), (2), and (3); generalized one-dimensional motion and applications of the higher order motion derivatives were not discussed. The video derivations of Eqs. (1) and (3) that I developed, noted earlier, were utilized as a post-class student resource. My colleague observed positive student response to this initial low-stakes pilot and plans to further develop this approach the next time she has the opportunity to teach calculus 2.

In the event of favorable further iterations, she will explore opportunities to disseminate ideas and resources to the mathematics teaching community.

Perhaps someone who comes across this paper many years in the future will read the first sentence and respond "yes, of course, I learned that in calculus, along with everyone else!" In the meantime, I hope that the material I have presented will be interesting to all *TPT* readers, and that the suggestions and resources I have outlined will be useful to those who would like to try incorporating Taylor series kinematics into their own formal or informal teaching, or who would like to explore collaborations with their mathematics colleagues. In this connection I note that high school instructors who regularly teach both AP Physics C and AP Calculus BC may have unique and expanded opportunities for impact.

**Acknowledgements**

I thank the following past and present Merrimack College colleagues for inspiration, feedback, and support relating to this project. I am grateful to Dan Tambasco, Emeritus Associate Professor of Physics, for pointing out to me many years ago that Eqs. (1), (2), and (3) are all Taylor series expansions and for reading through several early manuscript drafts. I thank Laura Hall-Seelig, Associate Professor of Mathematics, for discussions and feedback that informed this paper and the accompanying videos, and for undertaking an initial pilot of Taylor series kinematics in her calculus 2 course. I thank Chris Duston, Associate Professor of Physics, for reading through multiple drafts and providing helpful comments. Finally, I thank Sean Wright, Learning Experience Designer and Teaching Studio Manager at Merrimack's Center for Teaching and Learning Innovation, Instruction and Design, for carrying out important post-production video edits.

have concluded that introducing the concept of the jerk will help avoid the problem of students becoming mired in problems with constant acceleration … Carolyn Sumners, of the Huston Museum of Science … had quite a few activities that let students observe the jerk in amusement park rides … I would advocate that physics teachers devote some time to teaching the concept of the jerk to their students."